# Ultralow-voltage operation of light-emitting diodes

Yaxiao Lian[1#], Dongchen Lan[2#], Shiyu Xing[1#], Bingbing Guo[1], Runchen Lai[1], Baodan Zhao[1,3], Richard H. Friend[3], Dawei Di[1,3*]

1. State Key Laboratory of Modern Optical Instrumentation, College of Optical Science and Engineering; International Research Center for Advanced Photonics, Zhejiang University, Hangzhou, 310027, China
2. Australian Centre for Advanced Photovoltaics, University of New South Wales, Sydney, 2052, Australia
3. Cavendish Laboratory, University of Cambridge, JJ Thomson Avenue, Cambridge, CB3 0HE, United Kingdom

[#] These authors contributed equally.
*Corresponding author. E-mail: daweidi@zju.edu.cn

**The radiative recombination of injected charge carriers gives rise to electroluminescence (EL), a central process for light-emitting diode (LED) operation. It is often presumed in some emerging fields of optoelectronics, including perovskite and organic LEDs, that the minimum voltage required for light emission is the semiconductor bandgap divided by the elementary charge. Here we show for many classes of LEDs, including those based on metal halide perovskite, organic, chalcogenide quantum-dot and commercial III-V semiconductors, photon emission can be generally observed at record-low driving voltages of 36%-60% of their bandgaps, corresponding to a large apparent energy gain of 0.6-1.4 eV per emitted photon. Importantly, for various classes of LEDs with very different modes of charge injection and recombination (dark saturation current densities ranging from $\sim 10^{-35}$ to $\sim 10^{-21}$ mA cm$^{-2}$), their EL intensity-voltage curves under low voltages exhibit similar behaviors, revealing a universal origin of ultralow-voltage device operation. Finally, we demonstrate as a proof-of-concept that perovskite LEDs can transmit data efficiently to a silicon detector at 1V, a voltage below the silicon bandgap. Our work provides a fresh insight into the operational limits of electroluminescent diodes, highlighting the significant potential of integrating low-voltage LEDs with silicon electronics for next-generation communications and computational applications.**

The development of LEDs[1-6] has created far-reaching impacts on lighting, display and information industries. Emerging LED technologies, including organic LEDs (OLEDs)[4,7,8], quantum-dot LEDs (QLEDs)[9-11], and perovskite LEDs (PeLEDs)[12-18], are gaining significant attention due to their promise as next-generation light sources. The key mechanism responsible for the light emission from LEDs is electroluminescence (EL), the radiative recombination of injected electrons and holes under an external voltage. It has been suggested that the minimum (threshold) driving voltage required to create photons from the EL process equals to the bandgap ($E_g$) of the emissive material divided by the elementary charge ($q$), in consideration of the energy conservation principle[19,20], while free energy considerations allow a marginal energy gain of a few $kT$ (where $k$ is the Boltzmann constant and $T$ is the temperature)[21]. This minimum voltage may be reduced by invoking various mechanisms, including thermally-assisted upconversion[22-25], sequential charge injection[26], interfacial dipoles[27], triplet-triplet annihilation upconversion (TTA-UC)[28], and Auger processes[10,28-32]. Recently, an operating voltage of as low as ~70% of the nominal bandgaps was observed for LEDs based on III-V semiconductors, and this was attributed to enhanced radiative recombination enabled by a novel quantum well design[33]. For OLEDs, a minimum operating voltage of $0.45E_g/q$ was reported[30,32,34], though whether a TTA process could be used to explain the origin of this low operating voltage is still a subject of debate[32,34,35]. Sub-bandgap operating voltages were also observed for perovskite[16], and quantum-dot LEDs[10,11,31,36,37] (Supplementary Table 1). These observations lead to an open question of what the lowest possible driving voltages really are for electroluminescence, and whether they stem from the same origin. Ultralow-voltage operation of electroluminescence may contribute to the development of next-generation energy-efficient optoelectronics.

In this work, we began our investigation by measuring the minimum operating voltages of LEDs based on emerging materials systems. Perovskite LEDs were our first experimental subjects. We fabricated, without optimization, iodine-based near-infrared "NFPI"[38] and bromide-based green-emitting "PCPB"[39] perovskite LEDs with EQEs of ~10% (Fig. 1a&b and Extended Data Fig. 1a&b. See Methods for fabrication details). We observed that, for these perovskite LEDs, the minimum voltages for EL were 1.3 V and 1.9 V (Fig. 1a and Fig. 1b), while the EL peak photon energies were 1.56 eV and 2.4 eV (Extended Data Fig. 2), respectively (the minimum detectable photon flux is ~$10^{16}$ s$^{-1}$ m$^{-2}$ for our standard measurement setup, see Methods for details). Here, EL peak photon energies are used to provide conservative estimates of the bandgaps. The observed minimum operating voltages are 81% and 79% of the bandgaps for NFPI and PCPB perovskite LEDs, respectively. The observation of near- or sub-bandgap operating voltages for these LEDs is consistent with recent findings for efficient perovskite LEDs[15,16,40-42]. Further, we performed similar experiment for small-molecule OLEDs based on rubrene, polymer OLEDs based on F8BT, and II-VI chalcogenide QLEDs based on CdSe/ZnS quantum dots. Sub-bandgap voltage EL was similarly observed (Fig. 1c-e), in line with reports on these classes of LEDs[11,16,30,32,36,37,41,43,44].

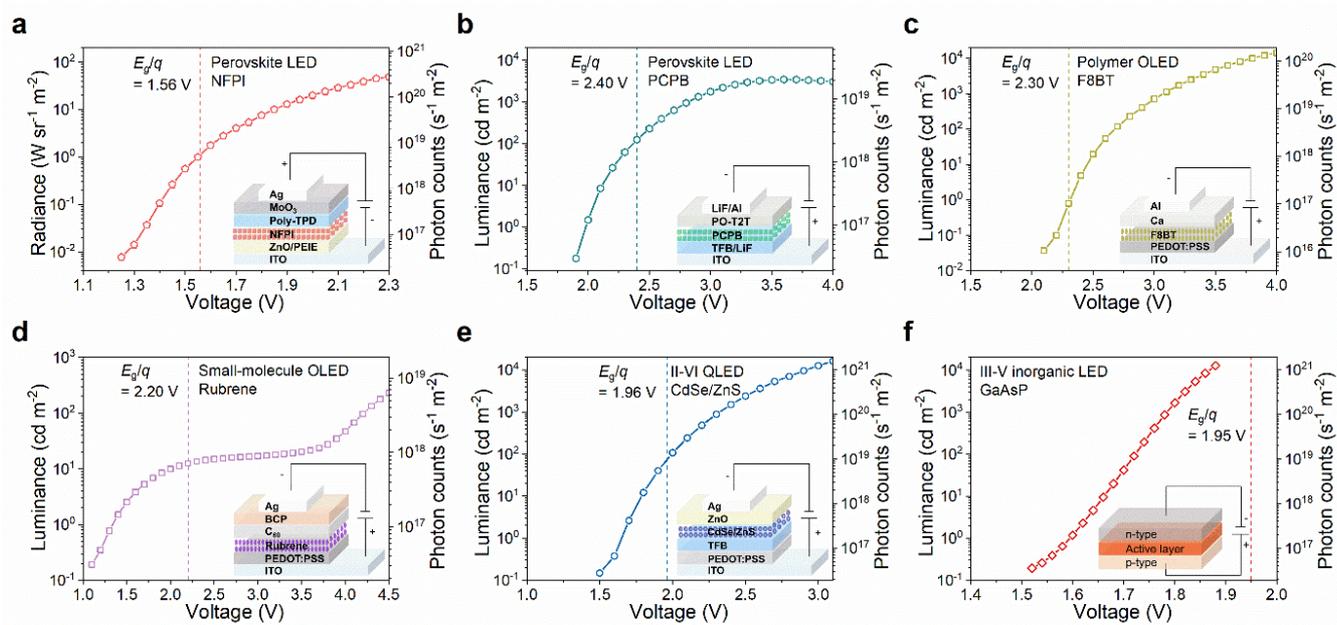

**Fig. 1 | EL intensity-voltage characteristics of different classes of LEDs. a**, Near-IR-emitting NFPI perovskite LED. **b**, Green PCPB perovskite LED. **c**, Polymer OLED based on F8BT. **d**, Small-molecule OLED based on rubrene. **e**, II-VI QLED based on CdSe/ZnS QDs. **f**, Commercial III-V inorganic LED based on GaAsP. The bandgaps for each emissive material are marked by dashed lines. Insets are schematics of the respective LED device structures.

Further, we measured commercial III-V LEDs based on GaAsP, GaP and AlGaP. Sub-bandgap voltage operation was similarly observed. For GaAsP-based LEDs with an $E_g$ of 1.95 eV, EL could be clearly observed under an applied voltage of 1.45 V ($0.74E_g$) using the same measurement setup (Methods). Importantly, the EL spectra remained unshifted as the driving voltages varied from above to clearly below the bandgaps (Fig. 2 and Extended Data Fig. 3 & 4), while sub-bandgap voltage EL is shown to be a general phenomenon.

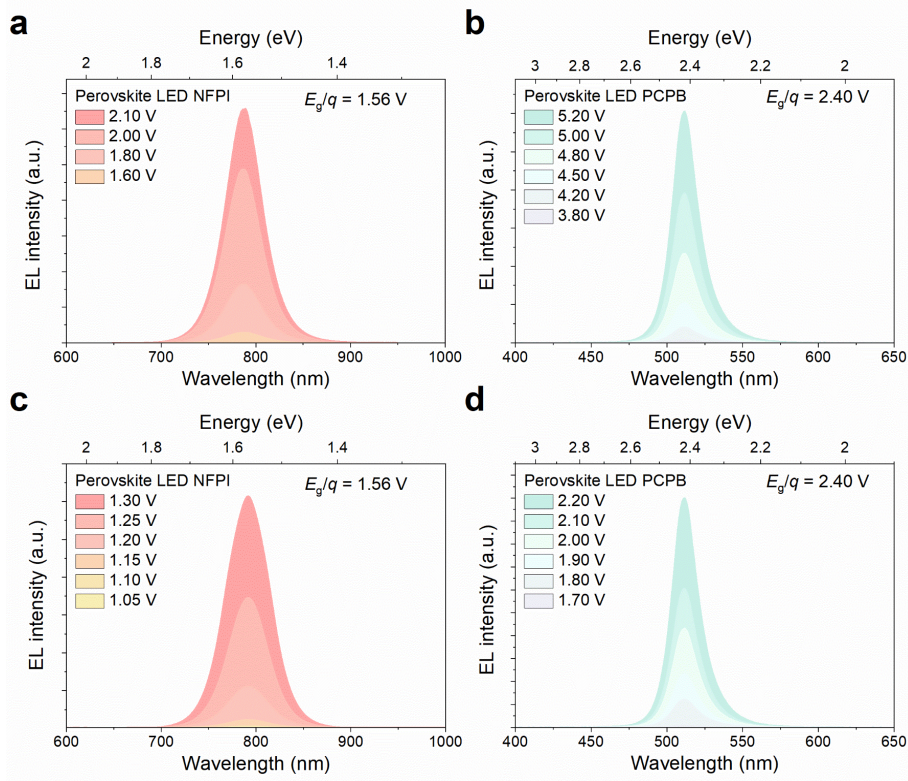

**Fig. 2 | EL spectra of perovskite LEDs under above- and sub-bandgap voltages. a**. **b**, EL spectra of NFPI and PCPB perovskite LEDs driven at different bias above $E_g/q$. **c**. **d**, EL spectra of NFPI and PCPB perovskite LEDs driven at sub-bandgap voltages, respectively.

A key scientific question is what the minimum voltages really are for the operation of LEDs. To find an answer to this problem, we employed a sensitive photon detection system (Extended Data Fig. 5) for the determination of the onset of EL, greatly enhancing the measurement sensitivity for weak photon emission (minimum detectable photon flux is ~$10^9$ s$^{-1}$ m$^{-2}$, which is 7 orders of magnitude more sensitive than the standard measurement setup; see Methods for details). EL was detected from our perovskite LEDs at voltages equivalent to 0.4-0.6$E_g$ (Fig. 3a), representing the lowest driving voltages reported for PeLEDs to date. For NFPI and PCPB perovskite LEDs emitting at ~790 nm and ~511 nm, the minimum voltages for observing EL were ~0.72 V and ~1.52 V, corresponding to $qV_m/E_g$ of ~46% and ~63% respectively (Fig. 3a). Here, $V_m$ denotes the measured minimum voltage required for generating detectable EL. The apparent energy gap $\Delta E = E_g - qV_m$, was as large as 0.9 eV. This is more than an order of magnitude greater than the room-temperature thermal energy ($kT = 26$ meV at 300 K).

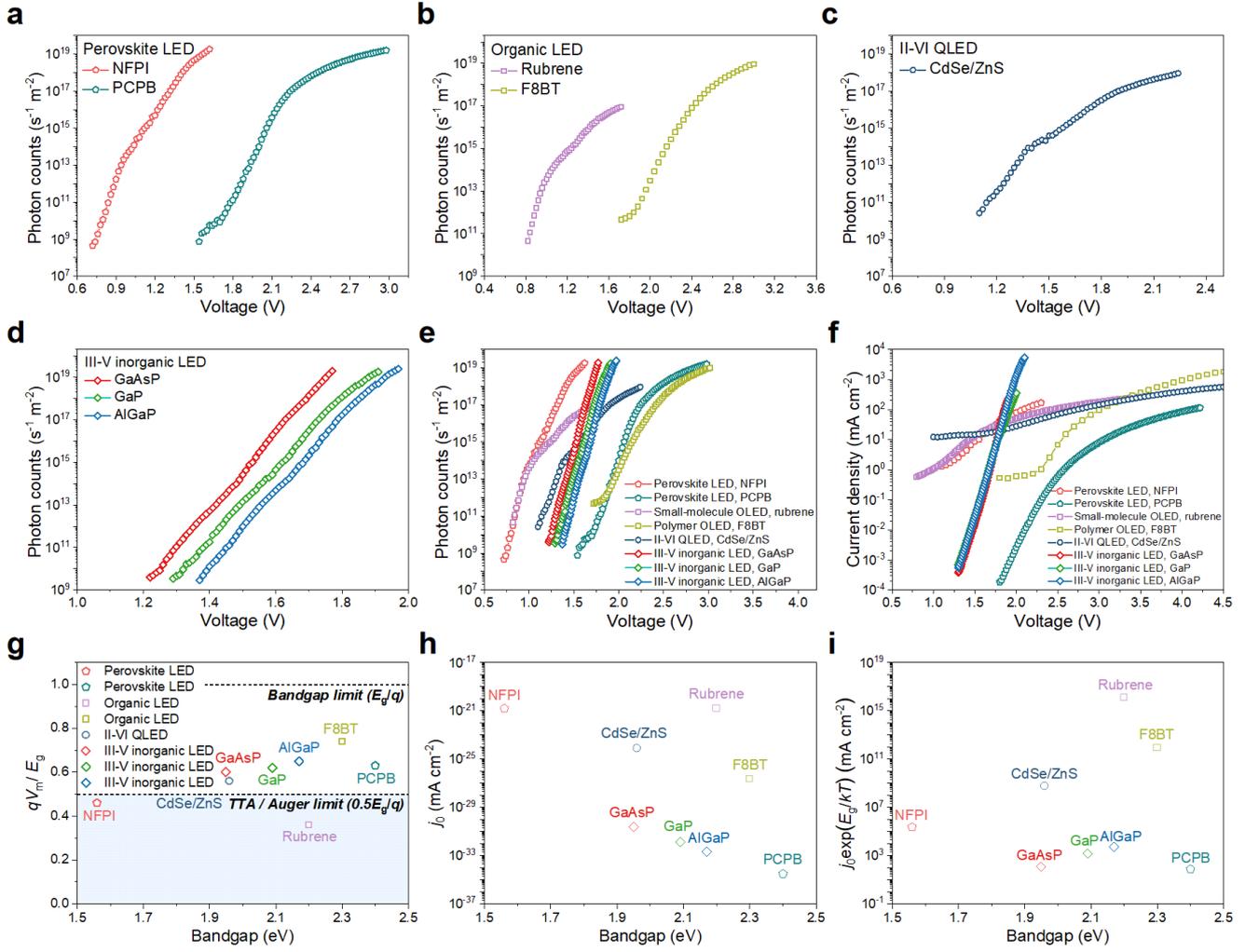

**Fig. 3 | EL intensity-voltage characteristics at near- and sub-bandgap voltages for different LEDs**. **a**, Perovskite LEDs with NFPI and PCPB emissive layers. **b**, Organic LEDs with rubrene and F8BT emissive layers. **c**, II-VI QLED based on CdSe/ZnS QDs. **d**, III-V inorganic LEDs based on GaAsP, GaP, and AlGaP. **e**, collection of EL intensity-voltage curves for different classes of LEDs in the same panel. **f**, Current density-voltage curves of different classes of LEDs. **g**, Measured $qV_m/E_g$ of different classes of LEDs. **h**, Dark saturation current densities ($j_0$) of different classes of LEDs. **i**, $j_0\exp(E_g/kT)$ of different classes of LEDs.

We similarly observed minimum operating voltages of 0.8 V ($0.36E_g/q$), 1.6 V ($0.74E_g/q$), 1.1 V ($0.56E_g/q$) and 1.25 V ($0.6E_g/q$) for small-molecule OLEDs based on rubrene, polymer OLEDs based on F8BT, II-VI QLEDs based on CdSe/ZnS QDs, and inorganic LEDs based on GaAsP, respectively (Fig. 3b-d). Record-low operating voltages were found for each classes of LEDs (Fig. 3e). We noted that the apparent energy gaps (Δ$E$) were on the order of ~1 eV. A summary of measured minimum voltages and Δ$E$ values we observed is provided in Supplementary Table 2, and the measurements were reproducible across a number of devices (Extended Data Fig. 6). Our experiments collectively demonstrate that the EL operation at sub-bandgap voltages is a universal phenomenon across different classes of LEDs (Extended Data Fig. 7a and Supplementary Table 2), and the operating voltages can reach values of ~$0.5E_g/q$ or below. For NPFI perovskite and rubrene devices, the measured $V_m$ values are even below the threshold voltage limits set by the TTA/Auger[10,28-32] processes (Fig. 3g). We note that these values may be reduced further by improving the instrumental sensitivity.

The current-voltage curves of various classes of LEDs show very different characteristics (Fig. 3f), which we consider is due to significant unipolar charge transport in some of the device with non-ideal device operation. However, all these diodes show remarkably similar EL intensity-voltage behaviours under low operating voltages (Fig. 3e), and follow the conventional diode equation described below as Eq. (1), where the light emission tracks the current density, $j(V)$:

$$j = j_0[\exp\left(\frac{qV}{nkT}\right) - 1] \quad (1)$$

where $j_0$ is the diode dark saturation current density ($j_0$ is negatively correlated with $E_g$ in a general form of $j_0 = \Lambda e^{-\frac{E_g}{kT}}$, where $\Lambda$ is related to materials properties. See Supplementary Note 1 for details), $n$ is the ideality factor, $k$ is the Boltzmann constant, $T$ is the temperature, and $V$ is the external voltage applied across the diode with mimimum influence from series resistance ($R_s$) at low voltages. While Eq. (1) is generally derived for unipolar p-n junction diodes, we see here that it clearly models the electron-hole recombination current for these diodes irrespective of the class of semiconductor and the choice of charge injection electrodes. In essence, the EL intensity ($I_{EL}$) is linked to the current density via the external quantum efficiency (EQE) of the LED:

$$I_{EL} = \text{EQE}(V)j(V, R_s)/q \quad (2)$$

where both EQE and $j$ are voltage-dependent, with $j$ depending also on $R_s$ at higher voltages.

Then the relation between the EL intensity and the applied voltage can be described by (See Supplementary Note 2 for details):

$$\log(I_{EL}) = \frac{\log(e)q}{nkT}V + \log[\text{EQE}(V)j_0] - \log(e)\,\text{W}[\frac{qR_sj_0}{nkT}\exp\left(\frac{qV}{nkT}\right)] \quad (3)$$

in which W is the Lambert W-function[45]. The EL intensity-voltage characteristics of our LEDs could be nicely fitted by this preliminary model (Extended Data Fig. 7).

We note that the dark saturation current density ($j_0$) varies greatly across different classes of LEDs, from ~$10^{-35}$ mA cm$^{-2}$ for PCPB perovskite, ~$10^{-31}$ mA cm$^{-2}$ for GaAsP, ~$10^{-27}$ mA cm$^{-2}$ for F8BT polymer to ~$10^{-21}$ mA cm$^{-2}$ for rubrene (Fig. 3h and Supplementary Table 3). $j_0$ contains important information on the physics of charge recombination in LEDs, and it depends on $E_g$ and $\Lambda$ (Supplementary Note 2). To allow a clearer comparison across different material systems, we calculated the bandgap-weighted dark saturation current densities ($j_0 e^{\frac{E_g}{kT}}$) for the LEDs we measured (Fig. 3i). Interestingly, the 'weighted $j_0$' values of the perovskite LEDs based on NPFI and PCPB now approaches the regime for III-V semiconductor LEDs (Fig. 3i and Supplementary Table 3). It is worth noting that such models originally developed for conventional inorganic semiconductor diodes can also be used to describe the general behavior of emerging classes of LEDs with vastly different materials properties, pointing toward a universality in the operating principles of different LEDs.

Using Eq. (3) it is possible to understand how the emissive material properties and the device design influence the apparent threshold voltages for EL (Fig. 4). An interesting observation is that higher driving voltages are required to generate the same photon flux for emissive materials with larger $E_g$. Indeed, this offers an explanation for why the apparent threshold voltages are generally higher for wider-bandgap LEDs. Similarly, higher series resistance tends to increase the apparent threshold voltages especially under larger bias, when the fractional potential drop on the active material becomes smaller. Clearly, low voltage operation is improved with higher EQE and also with reduced barriers and series resistance for charge injection. The effects of the latter can be seen by reducing the resistance of the charge-transport layers. Using an NFPI-based perovskite LED as an example, simply by reducing the thickness of the hole-transport layer, Poly(N,N'-bis-4-butylphenyl-N,N'-bisphenyl)benzidine (poly-TPD), the apparent threshold voltages can be lowered from 2.4 V to 1.4 V (Extended Data Fig. 8a). This can be attributed to the reduction of the series resistance in the perovskite LED. Similarly, for the PCPB-based perovskite LED, by reducing the thickness of (1,3,5-Triazine-2,4,6-triyl)tris(benzene-3,1-diyl)tris(diphenylphosphine oxide) (PO-T2T, $\mu_e$ ~4.4×10$^{-3}$ cm$^2$ V$^{-1}$ s$^{-1}$), an electron-transport material with much higher electron mobility than commonly used electron-transport materials such as bathophenanthroline (Bphen, $\mu_e$ ~5.2×10$^{-4}$ cm$^2$ V$^{-1}$ s$^{-1}$) and 1,3,5-Tris(1-phenyl-1H-benzimidazol-2-yl)benzene (TPBi, $\mu_e$ ~3.3×10$^{-5}$ cm$^2$ V$^{-1}$ s$^{-1}$), the apparent threshold voltages could be markedly reduced (Extended Data Fig. 8b).

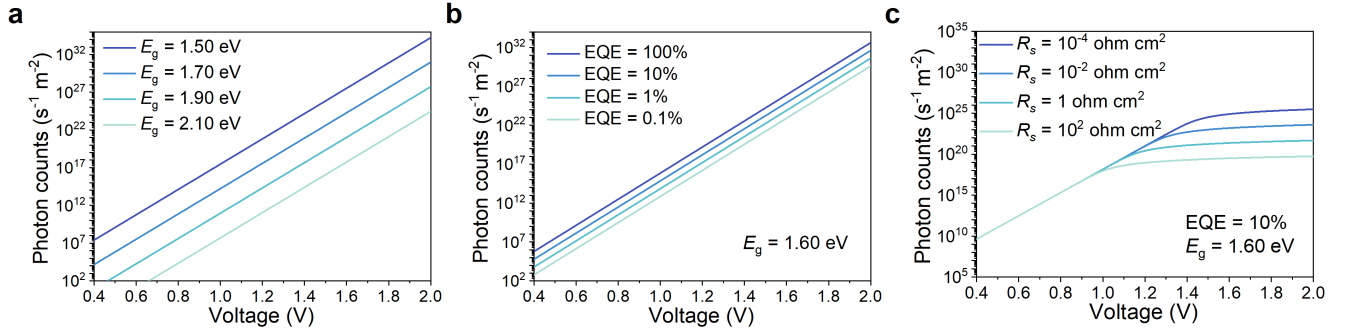

**Fig. 4 | EL intensity-voltage curves generated by a simple LED model. a**, Effect of bandgap. Ideal diodes with zero series resistance and an EQE of unity are assumed. **b**, Effect of EQE. A bandgap of 1.60 eV with zero series resistance is assumed. **c**, Effect of series resistance. A bandgap of 1.60 eV and an EQE of 10% are assumed.

A useful outcome of the low-voltage operation of LEDs is that these devices may be more versatile than conventional expectations. To provide an example of how this may provide benefits in practical applications, we employed our NFPI perovskite LEDs in a simple optical transmitter setup. With the application of sub-bandgap driving voltage of 1 V ($0.7E_g/q$), we were able to generate 1/0 (on/off) signals with a signal to noise ratio of 20 dB (Fig. 5b). The corresponding energy consumption for the LED to produce an optical pulse is as low as 140 pJ per bit for input frequencies ranging from 100 Hz to 1 MHz. The output pulse width is ~15 ns for an input pulse width of ~18 ns under the frequency range tested (Fig. 5b & Supplementary Table 4). Further reductions in energy consumption and pulse width may be possible by using a pulse generator with a smaller minimum pulse width. Remarkably, the voltage needed for optical data transimission (1V) is lower than the silicon bandgap (1.12 eV) divided by the elementary charge. Since it is theoretically possible to generate photons at voltages approaching zero, our results offer prospects for integrating low-cost perovskite LEDs with silicon-based devices for efficient optoelectronic logic operations, showing advantages over similar device integration using high-cost, epitaxially grown III-V semiconductors. More generally, as long as the bandgap of the emissive material is equal to or greater than that of silicon, it is possible to drive silicon-based devices with an LED operating at a voltage well below the silicon bandgap, showing great potential in logic and communications applications[46,47].

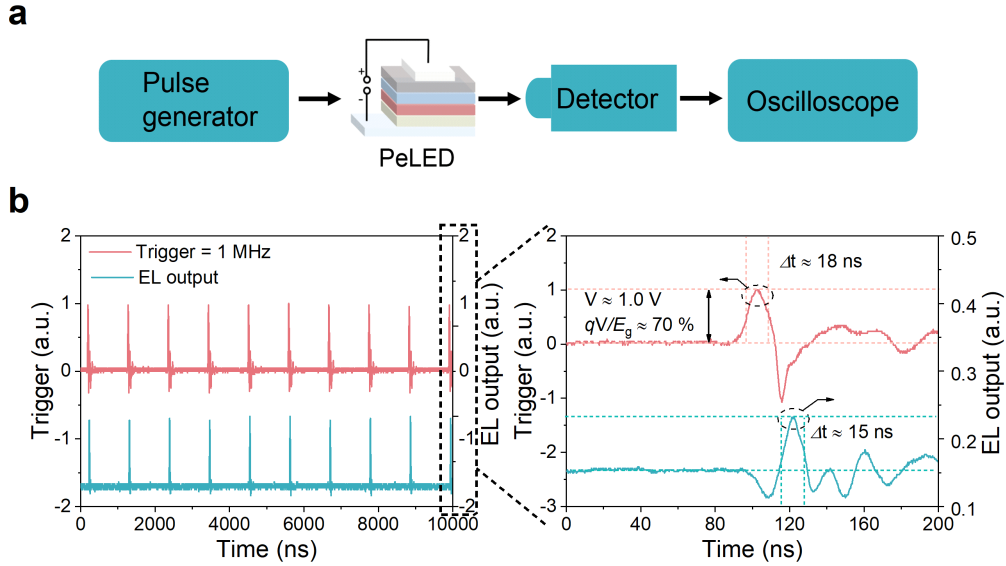

**Fig. 5 | Optical pulse generation from a perovskite LED under ultralow operating voltages. a**, Schematic diagram of a simple optical communications setup featuring an NFPI perovskite LED. **b**, The input electrical trigger from the pulse generator at a peak voltage of 1.0 V ($0.7E_g/q$), and the corresponding optical pulses (EL output) from the perovskite LEDs.

In summary, we have demonstrated, through high-sensitivity photon detection experiments, that the voltage for EL operation could reach values below 50% of the semiconductor bandgaps, exhibiting a large apparent energy gain of 0.6-1.4 eV per emitted photon. EL emission under ultralow voltages is a universal phenomenon across a broad range of LEDs based on perovskite, organic, II-VI chalcogenide quantum-dot and III-V semiconductors. Importantly, for these classes LEDs with very different modes of charge injection and recombination (e.g., dark saturation current densities ranging from ~$10^{-35}$ to ~$10^{-21}$ mA cm$^{-2}$), their EL intensity-voltage curves follow the conventional diode law, revealing the universality of ultralow-voltage device operation. Radiative recombination of non-thermal-equilibrium band-edge carrier populations under external bias is responsible for the low-voltage EL emission. We have demonstrated as a proof-of-concept that perovskite LEDs can transmit optical data efficiently to a silicon detector at voltages below the silicon bandgap, offering prospects for efficient data transmission at low costs. Beyond our efforts of understanding the operational limits of electroluminescence devices, these results may lead us to the under-explored territory of ultralow-voltage LEDs for communications, computational and energy applications.

**Methods**

**Materials**

Poly (9, 9-dioctylfluorene-co-N-(4-(3-methylpropyl)) diphenylamine) (TFB) was purchased from Luminescence Technology Corp. Poly [N, N'-bis(4-butylphenyl)-N,N'-bis(phenyl)-benzidine] (Poly-TPD) were purchased from American Dye Source. Colloidal CdSe/ZnS core-shell QDs were purchased from Guangdong Poly OptoElectronics Co., Ltd. Chlorobenzene (extra dry, 99.8%), octane (extra dry, >99%), ethanol (extra dry, 99.5%), N, N-dimethylformamide (DMF, 99.5%), Dimethyl sulfoxide (DMSO, HPLC grade) and ethyl acetate (HPLC grade) were purchased from J&K Chemical Ltd. PEDOT:PSS solution, 1-naphthylmethylamine iodide (NMAI, 99.9%), formamidinium iodide (FAI, 99.9%), PO-T2T (99.99%), $MoO_3$ (99.9%), 2-phenylethylammonium bromide (PEABr, 99.99%) and BCP (99.99%) was purchased from Xi'an Polymer Light Technology Corp. Caesium bromide (CsBr, 99.99%), lead bromide ($PbBr_2$, 99.999%), LiF (99.99%), $C_{60}$ (99.99%), tetraphenylphosphonium chloride (TPPCl), high-purity (99.99%) rubrene were purchased from Sigma-Aldrich. All materials were used as received without further purification.

**Fabrication of NFPI perovskite LEDs**

The perovskite precursor solution of 1-naphthylmethylamine iodide (NMAI), formamidinium iodide (FAI) and $PbI_2$ with a molar ratio of 2:1.8:2 dissolved in N, N-dimethylformamide (DMF) was prepared to form perovskite emissive layers with a composition of $NMA_2FA_{n-1}Pb_nI_{3n+1}$ (abbreviated as NFPI). The molar concentration for $PbI_2$ was 0.08 M.

The device structure of the NFPI PeLEDs was ITO/PEIE-coated ZnO/NFPI/Poly-TPD/$MoO_3$/Ag. Solutions of ZnO nanocrystals were spin coated onto the ITO-coated glass substrates at 5000 rpm for 60 s and annealed in air at 150 °C for 10 min. The substrates were transferred into a $N_2$ glovebox. Next, PEIE solution was spin-coated onto the ZnO surface at a speed of 5000 rpm for 60 s. The perovskite films were prepared by spin-coating the precursor solution onto the PEIE-treated ZnO films, followed by annealing at 100 °C for 10 min. Poly-TPD in chlorobenzene (12 mg·mL$^{-1}$) was spin-coated at 4000 rpm for 60 s. Finally, the $MoO_3$/Ag electrodes were deposited using a thermal evaporation system through a shadow mask under a base pressure of $4\times10^{-4}$ Pa. The device area was 5.25 mm$^{-2}$ as defined by the overlapping area of the ITO films and top electrodes. All the devices were encapsulated with UV epoxy (NOA81, Thorlabs)/cover glass before subsequent measurements.

**Fabrication of PCPB perovskite LEDs**

The perovskite precursor solution (molar ratio 5:5:2) was prepared by dissolving 110 mg lead bromide ($PbBr_2$), 64 mg caesium bromide (CsBr), and 24 mg 2-phenylethylammonium bromide (PEABr) in 1 mL dimethylsulfoxide (DMSO) and stirring overnight at room temperature, the additive is tetraphenylphosphonium chloride (TPPCl) and the volume ratio with perovskite precursor is 11:200. A quasi-2D/3D perovskite composition of $PEA_2Cs_{n-1}Pb_nBr_{3n+1}$ (abbreviated as PCPB) was expected to form.

The device structure of PCPB PeLEDs was ITO/TFB/LiF/PCPB/PO-T2T/LiF/Al. Pre-patterned ITO substrates (15 ohms/square) were cleaned using ultra-sonication in acetone and isopropanol for 15 min, respectively, and then dried with a nitrogen blow gun, after which the substrates were treated under UV-Ozone for 60 min. The ITO substrates were then transferred to a nitrogen-filled glovebox. TFB was spun-coated from chlorobenzene solution (6 mg mL$^{-1}$) at 5000 rpm and was annealed at 120 °C for 10 min. 1.3 nm of LiF was then evaporated at a pressure of $4 \times 10^{-4}$ Pa. Subsequently, the perovskite was spin-coated from the precursor solution at 5000 rpm to form a ~35 nm layer. Finally, PO-T2T (10 nm), LiF (0.8 nm) and Al (120 nm) were sequentially evaporated through a shadow mask under a base pressure of $4 \times 10^{-4}$ Pa. The device area was 5.25 mm$^{-2}$ as defined by the overlapping area of the ITO films and top electrodes. All the devices were encapsulated with UV epoxy (NOA81, Thorlabs)/cover glass before subsequent measurements. The deposition rate for thermal evaporation was calibrated and was kept at around 2 Å s$^{-1}$ during the evaporation process for materials except LiF, for which an evaporation rate of around 0.1 Å s$^{-1}$ was used.

**Fabrication of small-molecule OLEDs based on rubrene**

The device structure of rubrene-based OLEDs was ITO/PEDOT:PSS/rubrene/$C_{60}$/BCP/Ag. Organic materials were used as purchased without further purification. PEDOT:PSS was spin-coated on the substrate at 4000 rpm for 60 s, followed by annealing at 150 °C for 20 min. The thickness of the PEDOTS:PSS layer was around 40 nm. The PEDOT:PSS-coated ITO substrates were then transferred into the thermal evaporation system. A 35-nm thin layer of rubrene and a 25-nm thin layer of $C_{60}$ were deposited at a constant deposition rate of 0.5 Å s$^{-1}$. The substrate temperature was maintained at 80°C during deposition. Further deposition was done at room temperature. A 6-nm thin layer of BCP was deposited prior to the deposition of the top electrode. Devices were completed by evaporation of a 120-nm thin layer of Ag. Metal deposition was achieved through a shadow mask. The device area was 5.25 mm$^{-2}$ as defined by the overlapping area of the ITO films and top electrodes. All depositions were performed under a base pressure lower than $4\times10^{-4}$ Pa. The devices were encapsulated with UV epoxy (NOA81, Thorlabs)/cover glass before subsequent measurements.

**Fabrication of polymer OLEDs based on F8BT**

The device structure of F8BT-based polymer OLEDs was ITO/PEDOT:PSS/F8BT/Ca/Al. The PEDOT:PSS was spin-coated on the substrate at 7000 rpm for 60 s, followed by thermal annealing at 150 ˚C for 20 min. The thickness of the PEDOTS:PSS layer was around 30 nm. F8BT was deposited by spin-coating from solution (14 mg mL$^{-1}$ in chlorobenzene) at 5000 rpm, and annealed at 160 ˚C for 20 min, resulting in a film thickness of 50 nm. A 3.5 nm thin layer of Ca and 120 nm layer of Al were deposited by a thermal evaporation system under a base pressure lower than $4\times10^{-4}$ Pa. Metal deposition was achieved through a shadow mask. The device area was 5.25 mm$^{-2}$ as defined by the overlapping area of the ITO films and top electrodes. All the devices were encapsulated with UV epoxy (NOA81, Thorlabs)/cover glass before subsequent measurements.

**Fabrication of II-VI QLEDs based on CdSe/ZnS QDs**

The device structure of the II-VI QLEDs was ITO/PEDOT:PSS/TFB/QD/ZnO/Ag. The PEDOT:PSS layer was deposited by a two-step spin-coating process at 1000 rpm for 10s and 4000 rpm for 50 s, followed by annealing at 150 ˚C for 20 min. The PEDOT:PSS-coated substrates were transferred into a nitrogen-filled glovebox (O$_2$<1 ppm, H$_2$O <1 ppm) for subsequent processes. TFB was spin-coated from solution (in chlorobenzene, 12 mg mL$^{-1}$) at 2000 rpm for 60 s and baked at 150 ˚C for 20 min. CdSe/ZnS QD solution (in octane, ~15 mg mL$^{-1}$) and ZnO nanocrystals (in ethanol, ~30 mg mL$^{-1}$) were sequentially spin-coated onto the substrates at 2000 rpm for 60 s. Next, Ag electrodes (120 nm) were deposited by a thermal evaporation system under a base pressure of $<4\times10^{-4}$ Pa. The deposition of electrodes was achieved through a shadow mask. The active area of each device was 5.25 mm$^{-2}$ as defined by the overlapping area of the ITO films and top electrodes. The devices were encapsulated with UV epoxy (NOA81, Thorlabs)/cover glass before subsequent measurements.

**Characterization of LED performance**

Current density-voltage (J-V) characteristics were measured using a Keithley 2400 source-meter unit. The luminance and EQE data were obtained using an Everfine OLED-200 commercial LED performance analysis system. The photon flux and EL spectra were measured simultaneously using a charge-coupled device centred over the light-emitting pixel. The luminance (in cd m$^{-2}$) and radiance (in W sr$^{-1}$ m$^{-2}$) of the devices were calculated based on the angular distribution functions of the LED emission and the known spectral response of the charge-coupled device. This standard setup can measure EL reliably beyond a minimum photon flux of ~$1.4\times10^{15}$ s$^{-1}$ m$^{-2}$ sr$^{-1}$, which corresponds to a minimum detectable photon flux of ~$10^{16}$ s$^{-1}$ m$^{-2}$ for the LED devices. Additional EL spectra of the devices driven under near- and sub-bandgap voltages were collected by a fibre-coupled focus lens and measured using a QE Pro spectrometer (Ocean Optics).

**High-sensitivity photon detection experiments**

The measurement setup for the high-sensitivity photon detection experiments is illustrated in Extended Data Fig. 5a. Photons emitted from different classes of LEDs under near-and sub-bandgap voltages were detected by a Si-based single-photon avalanche photodiode (APD). The APD converts the photons from the LEDs into photocurrent, which is amplified by an amplifier. The photocurrent forms sharp pulses through a pulse shaper. These pulses are transmitted effectively through a BNC wire with low signal distortion. The controller converts the pulses to photon counts before transmitting data to the computer. It has an instrumental response time of ~0.2 ns. For each measurement, the minimum counts on the APD is on the order of 1000 s$^{-1}$, corresponding to a minimum detectable photon flux of ~$10^9$ s$^{-1}$ m$^{-2}$ for the LED devices.

The raw EL intensity data collected using the high-sensitivity photon detection system are shown in Extended Data Fig. 5b-g. Due to the finite collection efficiency and the intrinsic saturation characteristics of the APD, the raw data counts do not follow a linear relationship with the actual photon counts from the EL of the LEDs. In this work, we used the EL intensity-voltage data obtained from the commercial LED measurement system to calibrate the photon count-voltage response of the high-sensitivity system by driving the same LED under an identical voltage range. To extend the measurement range and to avoid saturation of the APD, EL from the LEDs was attenuated by a neutral density filter before entering the APD. The transfer function g($x$) for the calibration of the high-sensitivity setup can be expressed by

$$APD\ counts = \mathrm{g}(Photon\ counts \times w)$$

where $w$ is the attenuation factor set by the neutral density filter and the collection efficiency of the optics. It is possible to calculate the actual photon counts from the raw data collected from the APD, according to

$$Photon\ counts = 1/w \times g^{-1}(APD\ counts)$$

where $g^{-1}(x)$ is the inverse function of $g(x)$.

**Data availability**

The data that support the finding of this study are available from the corresponding author upon reasonable request.

**Acknowledgements**

This work was supported by the National Key Research and Development Program of China (grant no. 2018YFB2200401), the National Natural Science Foundation of China (NSFC) (61975180, 62005243), Kun-Peng Programme of Zhejiang Province (D.D.), the Natural Science Foundation of Zhejiang Province (LR21F050003), the Fundamental Research Funds for the Central Universities (2019QNA5005, 2020QNA5002), Zhejiang University Education Foundation Global Partnership Fund, and the Engineering and Physical Sciences Research Council (EPSRC). We thank Prof. Xiaogang Peng (Zhejiang University), Prof. Yizheng Jin (Zhejiang University), Dr. Linsong Cui (University of Cambridge) and Prof. Neil Greenham (University of Cambridge) for discussions.

**Author contributions**

D.D. conceived the project and planned the study. Y.L. and S.X. fabricated and characterized the LEDs. Y.L. performed the high-sensitivity EL experiments. D.L. established the diode model and derived relevant equations. D.L. and S. X. carried out data fitting and analysis. B.G., R.L. and B.Z. contributed to experiments and analyses. D.D., Y.L., S.X. and D.L. wrote the manuscript. R.H.F. provided suggestions for substantial revisions. All authors contributed to the work and commented on the paper.

**Competing interests**

The authors declare no competing interests.

**Corresponding author**

Correspondence to: Dawei Di (daweidi@zju.edu.cn).

# Extended data figures

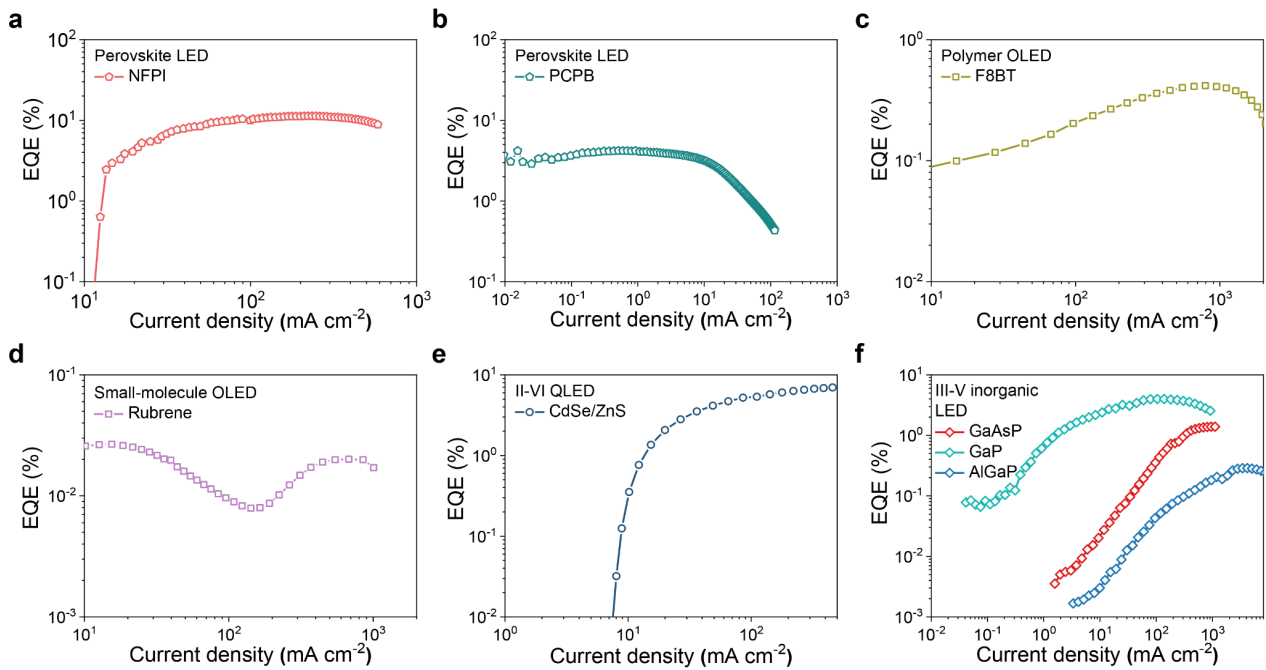

**Extended Data Fig. 1 | EQE versus current density curves for different classes of LEDs**. **a**, NIR-emitting NPFI perovskite LED. **b**, Green-emitting PCPB perovskite LED. **c,** Polymer OLED based on F8BT. **d**, Small-molecule OLED based on rubrene. **e**, II-VI QLED based on CdSe/ZnS QDs. **f**, Commercial III-V inorganic LEDs based on GaAsP, GaP and AlGaP.

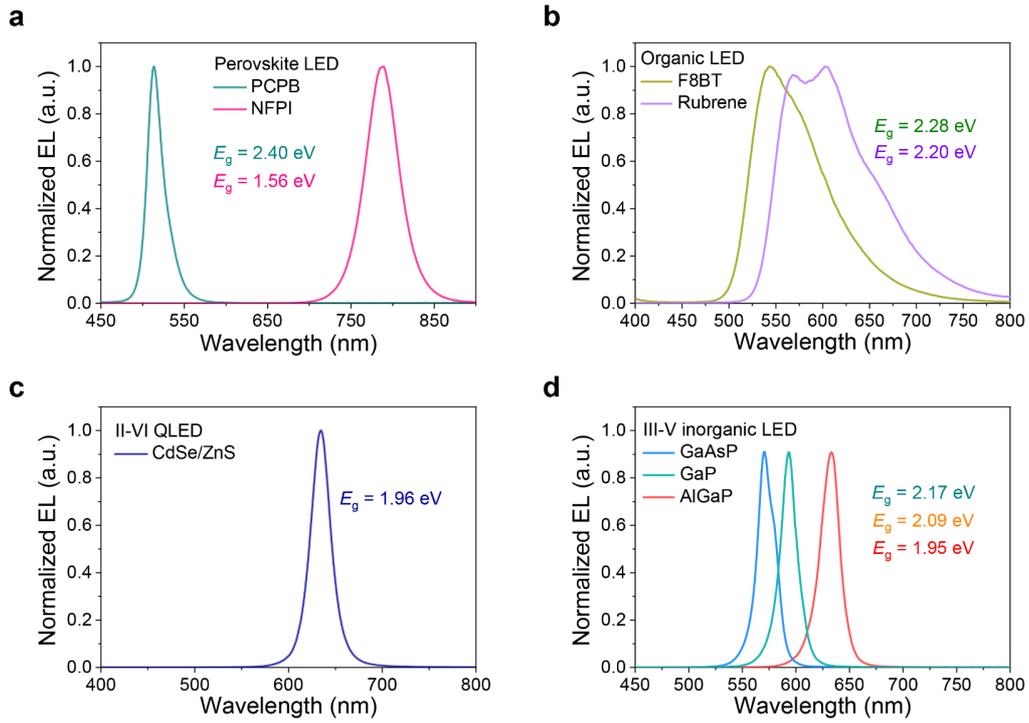

**Extended Data Fig. 2 | Normalized EL spectra of different LEDs. a**, NFPI and PCPB perovskite LEDs. **b**, OLEDs based on F8BT and rubrene. **c**, II-VI QLED based on CdSe/ZnS QDs. **d**, Commercial III-V inorganic LEDs.

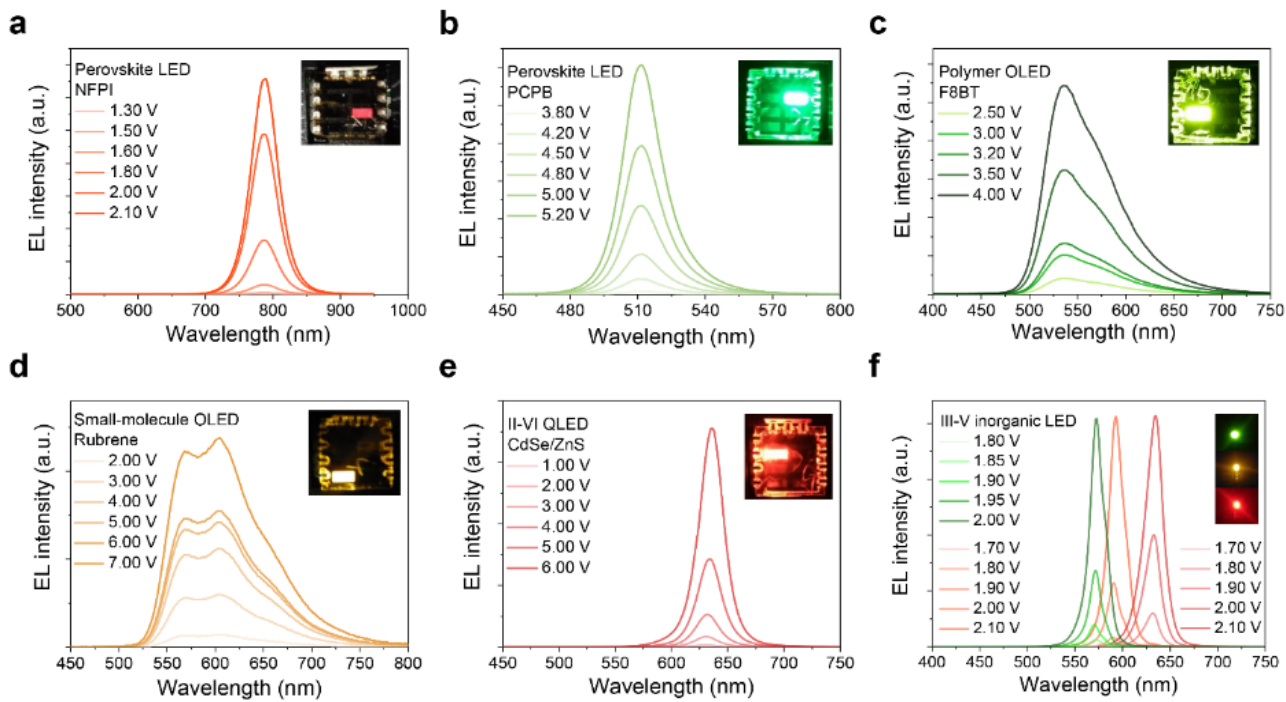

**Extended Data Fig. 3 | EL spectra of different classes of LEDs under different voltages. a**, Perovskite LED based on NFPI. **b**, Perovskite LED based on PCPB. **c**, Polymer OLED based on F8BT. **d**, Small-molecule OLED based on rubrene. **e**, II-VI QLED based on CdSe/ZnS QDs. **f**, Commercial III-V inorganic LEDs based on GaAsP, GaP and AlGaP.

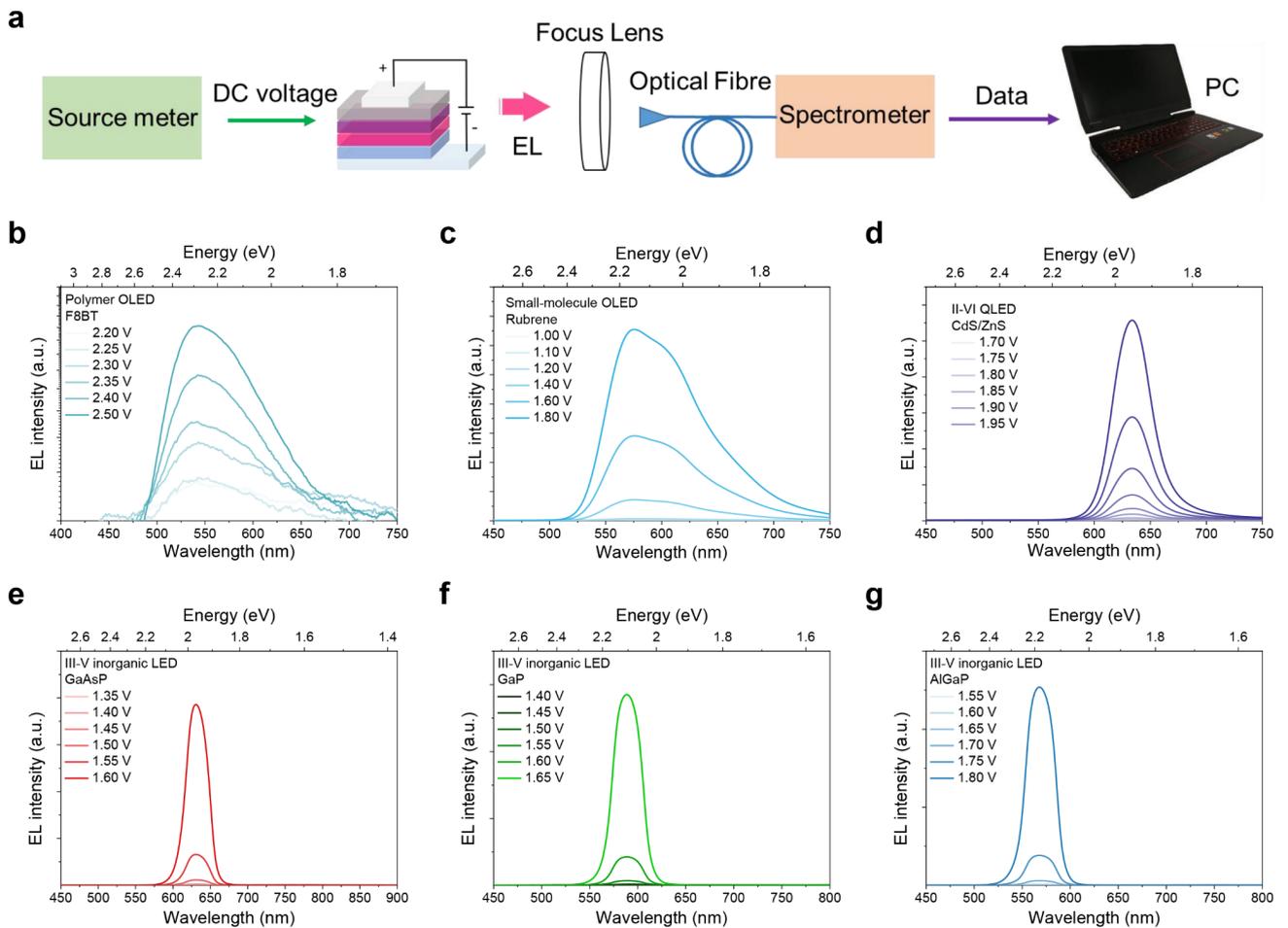

**Extended Data Fig. 4 | EL spectra of different classes of LEDs under sub-bandgap voltages. a**, Schematic of experimental setup for measuring EL spectra for LEDs driven at sub-bandgap voltages. **b**, Polymer OLED based on F8BT. **c**, Small-molecule OLED based on rubrene. **d**, II-VI QLED based on CdSe/ZnS QDs. **e-g**, Commercial III-V inorganic LEDs based on GaAsP, GaP, and AlGaP.

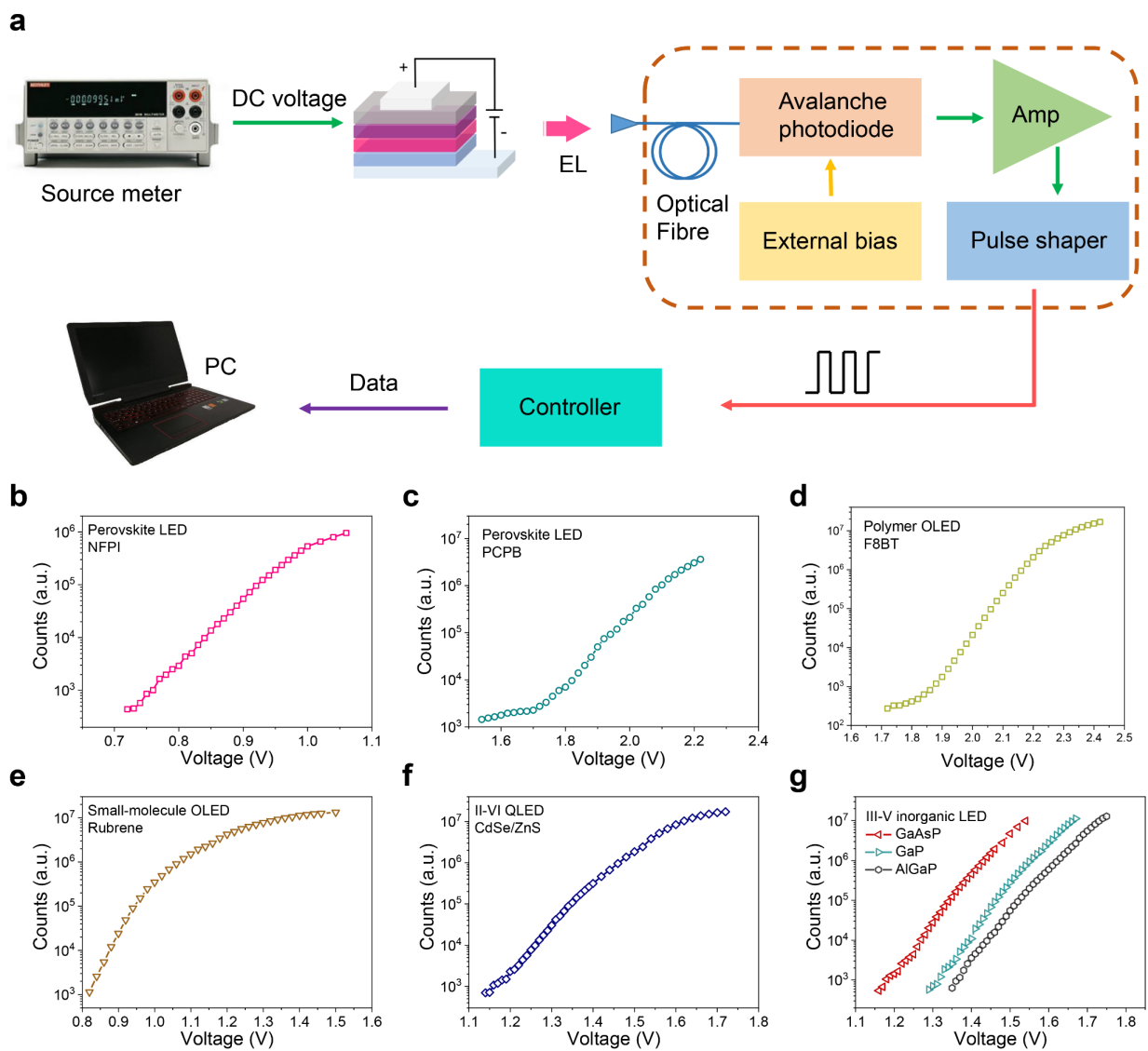

**Extended Data Fig. 5 | High-sensitivity photon detection experiments. a**, Schematic of the high-sensitivity photon detection system. The original APD counts-voltage characteristics of **b**, NFPI perovskite LED; **c**, PCPB perovskite LED; **d**, Polymer OLED based on F8BT; **e**, Small-molecule OLED based on rubrene; **f**, II-VI QLED based on CdSe/ZnS QDs; and **g**, Commercial III-V inorganic LEDs based on GaAsP, GaP and AlGaP.

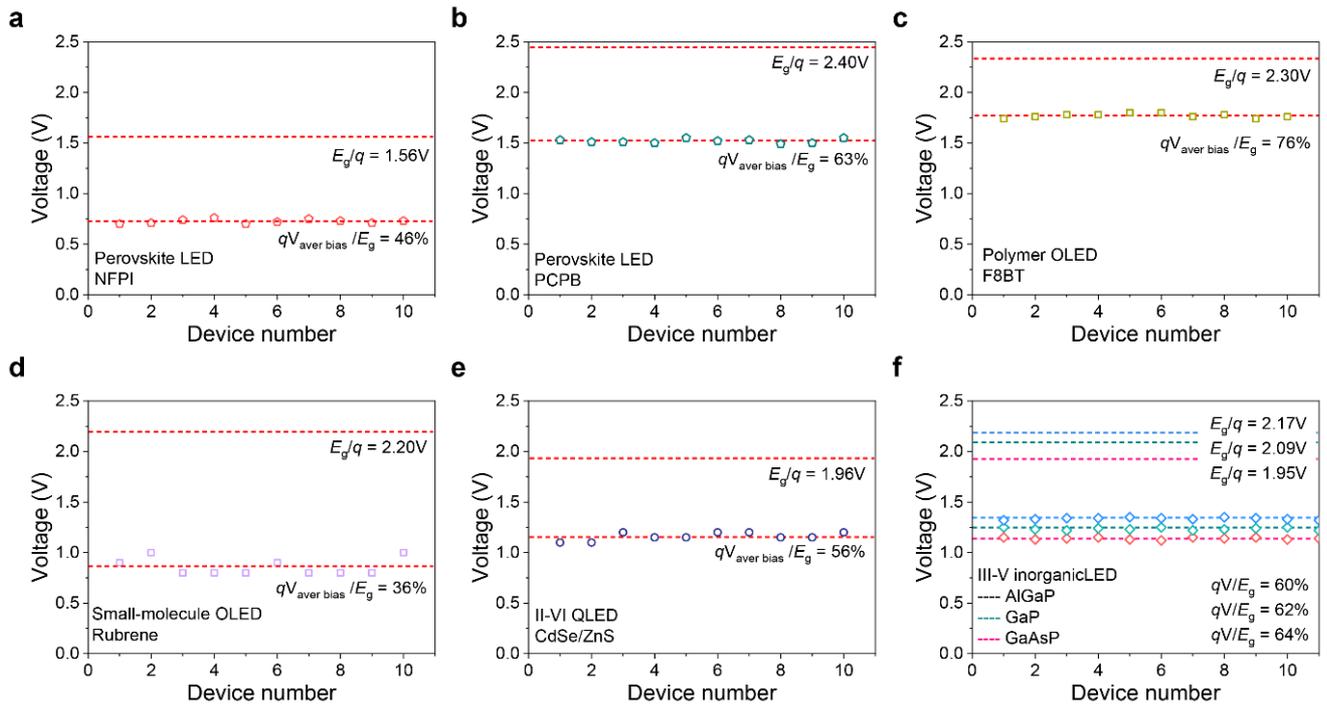

**Extended Data Fig. 6 | Statistics of the measured minimum operating voltages for different LEDs**. **a**, Perovskite LED based on NFPI. **b**, Perovskite LED based on PCPB. **c**, Polymer OLED based on F8BT. **d**, Small-molecule OLED based on rubrene. **e**, II-VI QLED based on CdSe/ZnS QDs. **f**, Commercial III-V inorganic LEDs based on GaAsP, GaP, and AlGaP.

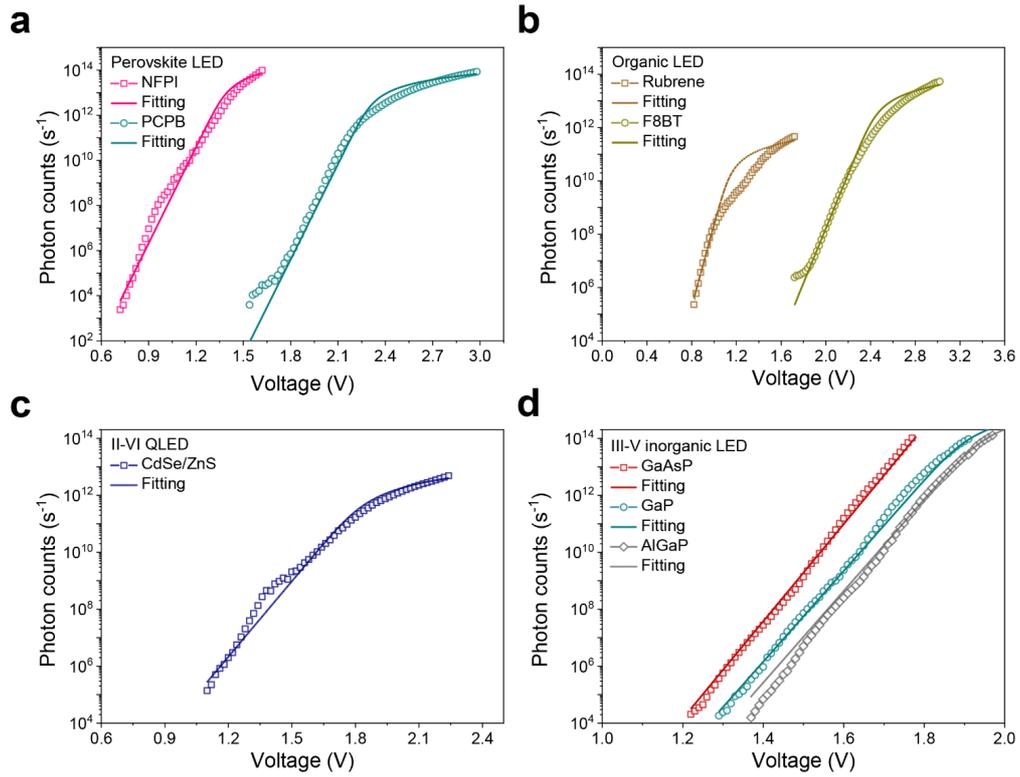

**Extended Data Fig. 7 | Measured EL intensity-voltage characteristics and curve fittings for different LEDs under near- and sub-bandgap voltages. a**, Perovskite LEDs. **b**, OLEDs. **c**, II-VI QLED. **d**, Commercial III-V inorganic LEDs.

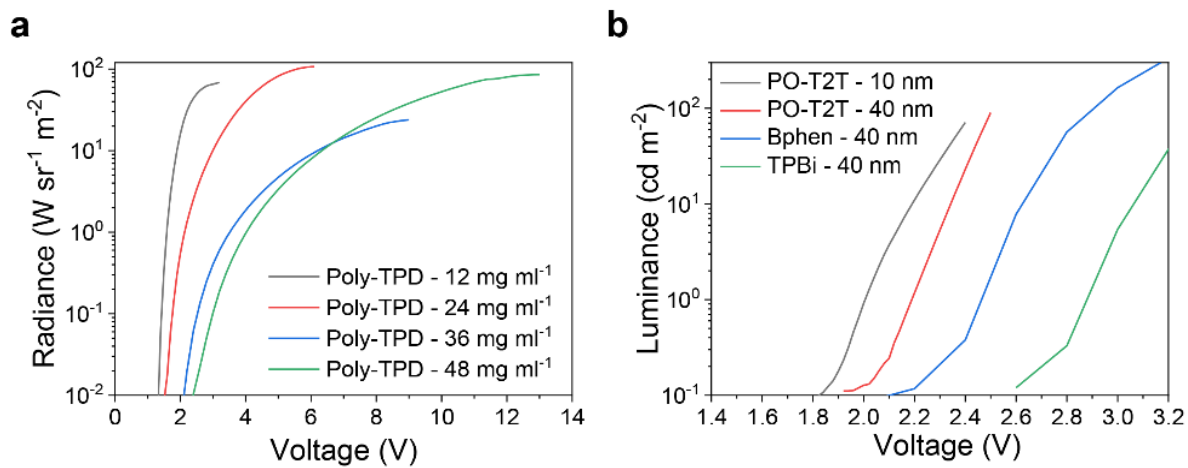

**Extended Data Fig. 8 | EL intensity-voltage characteristics of NFPI and PCPB perovskite LEDs. a**, Radiance-voltage characteristics of NFPI perovskite LEDs with various Poly-TPD thicknesses. **b**, Luminance-voltage characteristics of PCPB perovskite LEDs with different electron-transport layers.

Supplementary Information for:

**Ultralow-voltage operation of light-emitting diodes**

Yaxiao Lian[1#], Dongchen Lan[2#], Shiyu Xing[1#], Bingbing Guo[1], Runchen Lai[1], Baodan Zhao[1,3], Richard H. Friend[3], Dawei Di[1,3*]

1. State Key Laboratory of Modern Optical Instrumentation, College of Optical Science and Engineering; International Research Center for Advanced Photonics, Zhejiang University, Hangzhou, 310027, China

2. Australian Centre for Advanced Photovoltaics, University of New South Wales, Sydney, 2052, Australia

3. Cavendish Laboratory, University of Cambridge, JJ Thomson Avenue, Cambridge, CB3 0HE, United Kingdom

[#] These authors contributed equally.

*Corresponding author. E-mail: daweidi@zju.edu.cn

**Supplementary Table 1 | Measured minimum operating voltages of different LEDs from the literature.**

| Device type | Emissive material | Peak wavelength (nm) | Bandgap (eV) | Minimum voltage (V) | $qV_{in}/E_g$ | Ref |
|---|---|---|---|---|---|---|
| Perovskite LED | $(NMA)_2(FA)_{n-1}Pb_nI_{3n+1}$ | 795 | 1.56 | 1.25 | 80% | 16 |
| Perovskite LED | $(NMA)_2FAPb_2I_7$ | 786 | 1.58 | 1.5 | 95% | 37 |
| Perovskite LED | $CsPbBr_3$ | 525 | 2.36 | 1.9 | 81% | 39 |
| Polymer OLED | F8BT | 538 | 2.3 | 2.3 | 100% | 43 |
| Small-molecule OLED | Rubrene | 563 | 2.2 | 1.0 | 46% | 29 |
| II-VI QLED | CdSe/CdZnS | 600 | 2.1 | 1.7 | 81% | 10 |
| II-VI QLED | CdSe/CdS | 640 | 1.94 | 1.7 | 88% | 11 |
| III-V inorganic LED | $In_{0.2}Ga_{0.8}As$ | 953 | 1.3 | 0.9 | 70% | 32 |

**Supplementary Table 2 | Measured minimum operating voltages of different LEDs studied in this work.**

| Device type | Emissive material | Peak wavelength (nm) | Bandgap (eV) | Measured minimum voltage (V) | $\Delta E$ (eV) | $qV_m/E_g$ |
|---|---|---|---|---|---|---|
| Perovskite LED | NFPI perovskite | 790 | 1.56 | 0.72 | 0.9 | 46% |
| Perovskite LED | PCPB perovskite | 511 | 2.4 | 1.5 | 0.9 | 63% |
| Polymer OLED | F8BT | 538 | 2.3 | 1.7 | 0.6 | 74% |
| Small-molecule OLED | Rubrene | 563 | 2.2 | 0.8 | 1.4 | 36% |
| II-VI QLED | CdSe/ZnS | 631 | 1.96 | 1.1 | 0.86 | 56% |
| III-V inorganic LED | GaAsP | 570 | 2.17 | 1.25 | 0.92 | 60% |
| III-V inorganic LED | GaP | 593 | 2.09 | 1.3 | 0.79 | 62% |
| III-V inorganic LED | AlGaP | 633 | 1.95 | 1.4 | 0.55 | 65% |

**Supplementary Table 3 | Dark saturation current densities ($j_0$) and the corresponding $j_0 e^{\frac{E_g}{kT}}$ of different classes of LEDs.**

| Device type | Emissive material | Ideality factor $n$ | $j_0$ (mA cm$^{-2}$) | $j_0\exp(E_g/kT)$ (mA cm$^{-2}$) |
|---|---|---|---|---|
| Perovskite LED | NFPI perovskite | 1.30 | $1.6\times10^{-21}$ | $2.3\times10^{5}$ |
| Perovskite LED | PCPB perovskite | 1.22 | $3.1\times10^{-35}$ | $7.5\times10^{1}$ |
| Polymer OLED | F8BT | 1.65 | $2.4\times10^{-27}$ | $8.9\times10^{11}$ |
| Small-molecule OLED | Rubrene | 1.10 | $1.7\times10^{-21}$ | $1.3\times10^{16}$ |
| II-VI QLED | CdSe/ZnS | 1.32 | $8.2\times10^{-25}$ | $6.0\times10^{8}$ |
| III-V inorganic LED | GaAsP | 1.00 | $2.3\times10^{-31}$ | $1.2\times10^{2}$ |
| III-V inorganic LED | GaP | 1.03 | $1.3\times10^{-32}$ | $1.5\times10^{3}$ |
| III-V inorganic LED | AlGaP | 1.00 | $2.2\times10^{-33}$ | $5.4\times10^{3}$ |

**Supplementary Table 4 | Energy consumption of a perovskite LED working as a photon source in the optical transmitter setup.** The perovskite LED was driven by electrical pulses with a peak voltage of 1 V ($qV/E_g$ = 70%).

| Frequency (Hz) | Current (mA) | FWHM of EL pulses (ns) | Energy (pJ/bit) |
|---|---|---|---|
| 100 | 8 | 15.3 | 147 |
| 1,000 | 7.5 | 15.4 | 138 |
| 10,000 | 7.6 | 15.3 | 139 |
| 100,000 | 7.4 | 15.3 | 136 |
| 1,000,000 | 7.8 | 15.2 | 142 |

**Supplementary Note 1 | Further notes on $j_0$.**

As discussed in the main text, $j_0$ is a materials specific constant and is negatively correlated with $E_g$. For ideal diodes based on conventional inorganic semiconductors, $j_0$ can be described by the following equation[47].

$$j_0 = q n_i^2 B L \qquad (S1)$$

where $n_i$ is the intrinsic carrier concentration, $B$ is the radiative recombination constant, and $L$ is the thickness of the device active layer. Here, $n_i$ is related to $E_g$ according to the following equation.

$$n_i = \sqrt{N_V N_C} e^{-\frac{E_g}{2kT}} \qquad (S2)$$

where $N_V$, $N_C$ are the effective densities of states in the valence and conduction bands, respectively. The quantity B in Eq. (S1) can be described by[48]

$$B = \frac{2}{\tau_m N_M} \qquad (S3)$$

where $N_M$ is the majority carrier concentration and $\tau_m$ is the minority carrier lifetime. For simplicity, here the donor and acceptor concentrations on the two sides of the junction are assumed to be equal. Minority carrier electron and hole lifetimes, $\tau_e$ and $\tau_h$, are assumed to be equal to $\tau_m$. Substituting Eqs. (S2) and (S3) into Eq. (S1) gives:

$$j_0 = \frac{2}{\tau_m N_M} q L N_V N_C e^{-\frac{E_g}{kT}} \qquad (S4)$$

It can be seen that $j_0$ is negatively correlated with $E_g$. The actual values of $j_0$ are expected to differ greatly across different materials systems and devices, but it follows the general form below.

$$j_0 = \Lambda e^{-\frac{E_g}{kT}} \qquad (S5)$$

where $\Lambda$ is a quantity affected by materials properties and device design.

**Supplementary Note 2 | Derivation of the $I_{EL}$-$V$ relation.**

To accommodate both low- and moderate-voltage ranges where the effect of series resistance ($R_s$) can not be neglected, the current-voltage ($J$-$V$) characteristics of an LED can be described by

$$j = j_0(\exp[\frac{q(V-jR_s)}{nkT}] - 1) \tag{S6}$$

Inserting Eq. (S6) and into Eq. (2) gives the following:

$$\frac{qI_{EL}}{\text{EQE}(V)} = j_0 \left( e^{\frac{qV}{nkT} - \frac{q^2R_s}{nkT\text{EQE}(V)}I_{EL}} - 1 \right) \tag{S7}$$

Neglecting the minus-one term on the right-hand side, taking natural logarithm on both sides of the equation and rearranging the terms give

$$\ln\left(\frac{qI_{EL}}{\text{EQE}(V)j_0}\right) + \frac{q^2R_s}{nkT\text{EQE}(V)}I_{EL} = \frac{qV}{nkT} \tag{S8}$$

By inspection, Eq. (S8) takes the form

$$\ln(A + Bx) + Cx = D \tag{S9}$$

The Lambert W function gives the following solution:

$$x = -\frac{A}{B} + \frac{1}{C}W\left(\frac{C}{B}e^{\frac{AC}{B}+D}\right) \tag{S10}$$

Given $I_{EL}$ is the variable to be solved, Eq. (S8) is in the standard form of Eq. (S9). The solution for $I_{EL}$ is

$$I_{EL} = \frac{nkT\text{EQE}(V)}{q^2R_s} W\left(\frac{qR_sj_0}{nkT} e^{\frac{qV}{nkT}}\right) \tag{S11}$$

Taking the logarithm to Eq. (S11) and invoking $\log W(z) = \log(z) - W(z)\log(e)$ give the relation:

$$\log(I_{EL}) = \frac{q\log(e)}{nkT}V + \log\left(\frac{\text{EQE}(V)j_0}{q}\right) - \log(e)W\left(\frac{qR_sj_0}{nkT} e^{\frac{qV}{nkT}}\right) \tag{S12}$$

**Supplementary references**